\documentclass[lettersize,journal]{IEEEtran}
\usepackage{algorithm}
\usepackage{algorithmic}
\usepackage{booktabs}
\usepackage[switch]{lineno}
\usepackage{amsmath}
\usepackage{amssymb}
\usepackage[colorlinks,linkcolor=black,anchorcolor=black,citecolor=blue]{hyperref}
\usepackage{moreverb}
\usepackage{lipsum}
\usepackage{graphicx}
\usepackage{enumitem}
\usepackage{CJK}
\usepackage{graphicx}
\allowdisplaybreaks[4]
\usepackage{epstopdf}
\usepackage{cite}
\usepackage{subcaption}
\usepackage{array}
\usepackage{enumitem}
\usepackage{stfloats}
\usepackage{setspace}
\usepackage[T1]{fontenc}
\begin{document}
\title{Adaptive Cooperative Streaming of Holographic Video Over Wireless Networks: A Proximal Policy Optimization Solution}

\author{Wanli~Wen,
        Jiping~Yan,
        Yulu~Zhang,
        Zhen~Huang,
        Liang~Liang,
        and~Yunjian~Jia \vspace{-8mm}
\thanks{
The authors are with the School of Microelectronics and Communication Engineering, Chongqing University, Chongqing 400044, China (wanli\_wen@cqu.edu.cn, jiping\_yan@stu.cqu.edu.cn, yulu\_zhang@stu.cqu.edu.cn, ZhenHuang@cqu.edu.cn, liangliang@cqu.edu.cn, yunjian@cqu.edu.cn). \textit{(Corresponding author: Liang~Liang})}
}

\maketitle

\begin{abstract}
Adapting holographic video streaming to fluctuating wireless channels is essential to maintain consistent and satisfactory Quality of Experience (QoE) for users, which, however, is a challenging task due to the dynamic and uncertain characteristics of wireless networks. To address this issue, we propose a holographic video cooperative streaming framework designed for a generic wireless network in which multiple access points can cooperatively transmit video with different bitrates to multiple users. Additionally, we model a novel QoE metric tailored specifically for holographic video streaming, which can effectively encapsulate the nuances of holographic video quality, quality fluctuations, and rebuffering occurrences simultaneously. Furthermore, we formulate a formidable QoE maximization problem, which is a non-convex mixed integer nonlinear programming problem. Using proximal policy optimization (PPO), a new class of reinforcement learning algorithms, we devise a joint beamforming and bitrate control scheme, which can be wisely adapted to fluctuations in the wireless channel. The numerical results demonstrate the superiority of the proposed scheme over representative baselines.

\end{abstract}

\setlength{\textfloatsep}{0pt} 
\section{Introduction}

Holographic video represents a groundbreaking medium technology that is applicable in diverse fields such as entertainment, education, and healthcare. In contrast to panoramic video, which offers only three rotational degrees-of-freedom (DoF), holographic video delivers a 6-DoF experience, incorporating both rotational and translational movements. This expanded freedom further allows users to navigate freely within the virtual environment, providing a truly immersive Quality of Experience (QoE). However, streaming holographic video over existing wireless networks presents significant challenges compared to panoramic video.  First, streaming holographic video requires tens to hundreds of times more data than panoramic video \cite{liu2021pointcloud}, as it encodes a three-dimensional (3D) scene's complex light field information from multiple DoFs. Second, streaming holographic video must account for occlusions among holographic objects as well as the relative distances between objects and viewers \cite{joint_ICC, qoe_occlusion}, factors that substantially affect the user's QoE and add complexity to the wireless streaming process. Third, due to the dynamic nature of wireless networks, streaming holographic video should  intelligently adapt to fluctuating wireless channels to ensure consistent and satisfactory QoE for users. 

There has been a lot of research on adaptive video streaming, but most of them focus on panoramic video. For example, the authors in \cite{mec} investigated a 3-DoF panoramic video streaming system and proposed a dynamic communication resource allocation scheme using a reinforcement learning (RL) approach \cite{du2020machine}. In \cite{zhang2021joint}, a video bitrate control algorithm was proposed that can adapt to wireless channel, with the aim of ensuring satisfactory QoE in a live streaming scenario. The authors in \cite{guo2021power} studied energy-efficient wireless streaming of panoramic video by jointly optimizing the beamforming and bitrate control according to the user's personalized field of view (FoV). In \cite{two-tier}, the authors proposed a communication resource allocation algorithm based on traditional numerical approaches to maximize QoE in a panoramic video streaming system. However, the above streaming schemes are designed mainly for panoramic video, which may not be directly applicable for holographic video streaming.

Currently, adaptive streaming of holographic video has garnered the attention of a number of researchers. For example, the authors in \cite{3dtile,qoe_occlusion, joint_ICC, vtc} proposed a joint communication and computation resource allocation scheme \cite{qoe_occlusion, joint_ICC}, as well as a bitrate control scheme \cite{3dtile, vtc}, employing traditional numerical optimization methods \cite{3dtile, qoe_occlusion} and RL approaches \cite{joint_ICC, vtc}, with the aim of improving user QoE. Note that existing research efforts typically consider a single access point (AP) to support video transmission, potentially limiting the channel capacity to meet the high-rate demand for holographic video. To address this challenge, our previous work \cite{wenjoint} introduced a cooperative holographic video streaming framework. This framework allows multiple APs to collaborate to transmit video to users and has been proven to excel in video quality based on numerical results. However, the QoE model in \cite{wenjoint} only covers the influence of video quality, but overlooks other crucial aspects of video streaming, such as quality fluctuations and rebuffering. Furthermore, the streaming scheme in \cite{wenjoint} is achieved based on the traditional numerical optimization method, which suffers from high complexity and is difficult to adapt to rapidly fluctuating wireless channels. 

In this work, we aim to address the aforementioned issues and achieve three key contributions. First, we develop a framework for cooperative holographic video streaming over a generic wireless network, enabling multiple APs to simultaneously transmit videos at varying bitrates to multiple users. Second, we introduce a novel QoE metric specifically designed for holographic video streaming that effectively captures the intricacies of video quality, quality fluctuations, and rebuffering events.  Finally, we formulate a QoE maximization problem, which is a non-convex mixed integer nonlinear programming (MINLP) problem. By applying the Proximal Policy Optimization (PPO) algorithm, we devise a joint beamforming and bitrate control scheme that can dynamically adapt to changes in wireless channel conditions.

\section{System Model}
\subsection{Network Model}
\begin{figure}
    \centering
    \includegraphics[width=3.0in,height=1.8in]{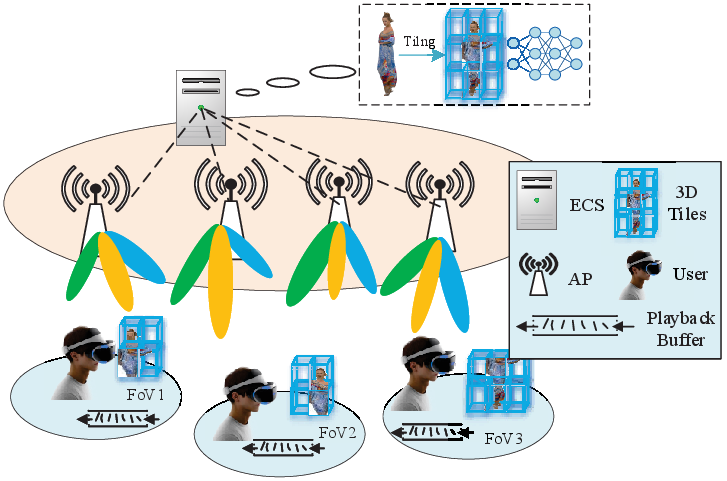}
    \caption{{\centering}System model.}
    \label{fig1}
\end{figure}

Let us consider a generic wireless edge network, as depicted in Fig.~\ref{fig1}, which consists of an edge cloud server (ECS), a set $\mathcal{M}\triangleq\{1,2,\cdots,M\}$ of $M$ APs, and a set $\mathcal{K}\triangleq\{1,2,\cdots,K\}$ of $K$ users.  The content server stores a holographic video and connects to AP $m\in\mathcal{M}$ via a high-speed and low-latency backhaul link, such as optical fiber. Through video playback device, user $k\in\mathcal{K}$ is streaming the holographic video from the ECS via the APs in $\mathcal{M}$. We consider a time-slotted system in which time is divided into slots indexed by $t$, each with a duration of $\tau$, and users request video at the beginning of each time slot.  We focus on the streaming operation during $T$ slots, which is represented by $\mathcal{T}\triangleq\{1,2,\ldots,T\}$.

To improve the transmission efficiency of the holographic video, we consider exploiting 3D tiling at the server side, where the holographic video is spatially partitioned into $N$ tiles \cite{3dtile}, represented by $\mathcal{N}\triangleq\{1,2,\ldots,N\}$. Note that limited by the human visual system, user $k$ can only see part of the tiles in $\mathcal{N}$. We use $\mathcal{N}_k(t)\in \mathcal{N}$ to denote the set of tiles within FoV of user $k$ at time $t$. Each tile is encoded into $L$ quality levels. Define $\mathcal{L}\triangleq\{1,2,\ldots,L\}$ to be the quality level set. A higher quality level corresponds to a larger bitrate. Let $\mu_l$ (in bps) indicate the tile's bitrate with quality level $l\in \mathcal{L}$, where $\mu _1\le \mu _2 \le \dots  \le \mu _L$ \cite{guo2021power} and the set of bitrates is denoted by $\mathcal{B}$. Define $l_{k,n}(t)$ to be the quality level of tile $n$ sent to user $k$ in slot $t$ which satisfies
\begin{align}\label{lkn}
    l_{k,n}(t) \in \mathcal{L},\;k\in \mathcal{K},\; n\in \mathcal{N},\;t\in \mathcal{T}.
\end{align}Let ${\mathbf l}\triangleq ({\mathbf l}(t))_{t\in \mathcal{T}}$ with ${\mathbf l}(t) \triangleq (l_{k,n}(t))_{n\in \mathcal{N},k\in \mathcal{K}}$ denote the bitrate selection design. For each tile, there exist both compressed and uncompressed bitrate versions. Let $\eta_{k,n}(t)$ represent the variable indicating whether the bitrate of tile $n$ requested by user $k$ at time $t$ is compressed ($\eta_{k,n}(t) = 1$) or uncompressed ($\eta_{k,n}(t) = 0$), i.e.,
\begin{align}\label{etakn}
    \eta_{k,n}(t)\in \{0,1\},\; k\in \mathcal{K},\; n\in \mathcal{N},\;t\in \mathcal{T}.
\end{align}For a compressed tile, we denote by $\varphi \in [0,1]$ its compression ratio. Let ${\boldsymbol \eta}\triangleq ({\boldsymbol \eta}(t))_{t\in \mathcal{T}}$ with ${\boldsymbol \eta}(t) \triangleq (\eta_{k,n}(t))_{n\in \mathcal{N},k\in \mathcal{K}}$ denote the compression design. %

We consider that each AP is equipped with $I$ transmit antennas and each user has a single receive antenna. Let $\mathbf{w}_{k,m}(t) \in \mathbb{C}^I$ denote the transmission beamformer on AP $n$ for user $k$ in slot $t$, which satisfies
\begin{align} \label{maxpower}
\sum_{k\in \mathcal{K}} \left\|\mathbf{w}_{k,m}(t)\right\|^2 \le P_m,\ m\in \mathcal{M},\ t\in \mathcal{T}.
\end{align}
Here, $P_m$ denotes the maximum transmit power of AP $m$. Let ${\mathbf w}\triangleq ({\mathbf w}(t))_{t\in \mathcal{T}}$ with ${\mathbf w}(t) \triangleq (\mathbf{w}_{k,m}(t))_{k\in \mathcal{K},m\in \mathcal{M}}$ denote the beamforming design. Denote by $\mathbf{h}_{k,m}(t)\in \mathbb{C}^I$ the wireless channel vector that exists between the $m$-th AP and the $k$-th user in slot $t$. We adopt a block fading model for the wireless channel, wherein the channel vector is assumed to be static within a given time slot but may vary from one slot to another \cite{zhang2021joint,guo2021power,two-tier}.\footnote{Note that the block fading model encapsulates the dynamic and uncertain characteristics of wireless networks, as it accounts for the variability of channel conditions over different slots.} Now, we can calculate the signal-to-interference-plus-noise ratio (SINR) for user $k$~as:
\begin{align}\label{eqSINR}
\gamma_k(t)=\frac{\left| {\sum\limits_{m\in \mathcal{M}} \mathbf{h}^{\rm H}_{k,m}(t)\mathbf{w}_{k,m}(t)}\right|^2}{\sum\limits_{k'\in \mathcal{K}\setminus \{k\}}^{} \left |\sum\limits_{m\in \mathcal{M}}  \mathbf{h}^{\rm H}_{k,m}(t)\mathbf{w}_{k',m}(t)  \right |^2   +N_0W},
\end{align}
where $N_0$ is the power spectral density of the noise, $W$ denotes the bandwidth of the wireless channel, and $(\cdot)^{\rm H}$ denotes the hermitian transpose operator. Based on (\ref{eqSINR}), the achievable data rate at user $k$ can be expressed as $r_k(t)={W}\log_2(1+\gamma_k(t))$. To ensure a reliable video transmission, we enforce the following constraint on $\gamma_k(t)$,~i.e.,
\begin{align}\label{sinr}
\gamma_k(t)\ge \xi, \  k \in \mathcal{K},\;t\in \mathcal{T},
\end{align} 
where $\xi$ is a SINR threshold. Let $T_g$ denote the play time of each tile. Then, the total transmission time of the tiles inside user $k$'s FoV in slot $t$ can be calculated as
\begin{align} \label{transmission_T}
T_k^{\rm r}(t)=\frac{1}{r_k(t)} \sum_{n\in \mathcal N_k(t)}^{}\Big(\big(1-\eta_{k,n}(t)\big) \mu _{l_{k,n}(t)}T_g\nonumber\\
+\   \varphi  \eta_{k,n}(t) \mu _{l_{k,n}(t)}T_g\Big).
\end{align}
Furthermore, when the received tile is a compressed version, it needs to be decoded depending on the computing capability of user device. The decoding time can be given by
\begin{align} \label{decode_T}
T_k^{\rm d}(t)=\frac{ \sum_{n\in \mathcal N_k(t)}^{}\varphi \eta_{k,n}(t) \mu _{l_{k,n}(t)}T_g} {C_{\rm max}\ b}.
\end{align}Here, $C_{\rm max}$ (in cycles per second) and $b$ (in bits per cycle) indicate the maximum number of CPU operations per second in each user's device, and the data volume the CPU can decode (decompress) in one cycle, respectively. Finally, the video transmission and decoding should be completed within slot $t$, so based on (\ref{transmission_T}) and (\ref{decode_T}), we have the following constraint:
\begin{align}\label{slot}
T_k^{\rm r}(t)+T_k^{\rm d}(t)\le\tau, \ k \in \mathcal{K},\;t\in \mathcal{T}.
\end{align}

\subsection{QoE Model}
In line with established research \cite{joint_ICC,zhang2021joint,vtc,wenjoint}, our QoE model integrates three  factors: video quality, video fluctuation, and rebuffering. The video quality is a subjective measure from the user's perspective that reflects the visual clarity. Video fluctuation refers to variations in quality over time, which can degrade the viewing experience. Rebuffering, characterized by playback interruptions due to depleted buffers, further detracts from QoE by disrupting the continuity of content. Our model will incorporate these aspects to deliver an enhanced and seamless holographic video streaming service. Specifically, the QoE model for user $k$ in slot $t$ is defined as\footnote{The compression ratio $\varphi$ indirectly affects the QoE through the constraint in (8), as a lower compression ratio reduces transmission time but may degrade video quality.}
\begin{align}\label{qoe}
{\rm QoE}_k(t) \triangleq Q_k(t) - \alpha_1 \Delta Q_k(t)  - \alpha_2 B_k(t),
\end{align}
where $B_k(t)$ represents the quality impairment caused by rebuffering event, and $\alpha_1>0$ and $\alpha_2>0$ are the weight factors. 
$Q_k(t)$, $\Delta Q_k(t)$, and $B_k(t)$ are explained as follows.
\begin{itemize}
    \item \textit{Video Quality}: This is closely tied to three crucial factors: video bitrate, the distance between the user and tiles, and the occlusion between tiles. Specifically, if a tile is invisible (e.g., outside the user’s FoV or completely occluded by another tile), then transmitting it would contribute nothing to the viewing experience. For a visible tile, on the other hand, the holographic video quality is usually considered as logarithmically proportional to the video bitrate \cite{wenjoint} and inversely proportional to the spatial distance between the user and the tile \cite{joint_ICC}. In addition, the video quality is inversely proportional to the tile's occlusion level \cite{qoe_occlusion}. Therefore, the holographic video quality of user $k$ in slot $t$ can be defined as
\begin{align*}
Q_{k}(t)=  \sum_{n\in \mathcal N_k(t)}^{} \left(\frac{w_{\rm dist}}{{\rm dist}_{k,n}(t)}+ \frac{w_o}{O_{k,n}(t)}\right) \ln_{}{\frac{w_{\mu} }{\mu _L}\mu_{l_{k,n}(t)} } .
\end{align*}Here, ${\rm dist}_{k,n}(t)$ is the fixed virtual distance between user's viewport and tile $n$, $O_{k,n}(t)$ is the occlusion level of the tile which reflects the tile's importance for users. In addition, $w_{\rm dist}$, ${w_o}$ and $w_\mu$ are the weighting factors.

\item \textit{Video Fluctuation}: As the adaptive video transmission mechanism leads to varying video quality, ensuring minimal fluctuations  is crucial for QoE \cite{zhang2021joint}. The change in video quality for user $k$ in slot $t$, compared to the previous slot $t-1$, is defined as
\begin{align*}
\Delta Q_k(t)=\left |  Q_{k}(t)-Q_{k}(t-1)\right |,
\end{align*}where the notation $|\cdot|$ means taking the absolute value. And we set $Q_{k}(0)=0$ for $k \in \mathcal{K}$.
\item \textit{Rebuffering}: A temporary pause in video playback due to network-induced buffering refills will result in rebuffering. We denote $G_k(t)$ as the buffer status of user $k$ in slot $t$. $G_k(t)$ is generally influenced by the buffer status at $t-1$, tile duration, and slot duration \cite{vtc}. It can be represented as
\begin{align*}
G_k(t)=\min(G_{\max},[G_k(t-1)+T_g-\tau]^+).
\end{align*}Here, $\left[x\right ]^+ \triangleq \max(x,0)$, where $G_{\max}$ signifies the upper limit of the buffer. $G_k(t) = 0$ indicates no available video to play, leading to rebuffering. We define $B_k(t)$ to represent the QoE degradation induced by rebuffering:
\begin{align*}
    B_k(t)= \textbf{1}_{\{G_k(t)=0\}}.
\end{align*}Here, $\textbf{1}_{\{G_k(t)=0\}}$ equals 1 if ${G_k(t)=0}$, and 0, otherwise.  
\end{itemize}

\section{Problem Formulation and Solution}
In this section, we start by formulating a QoE maximization problem and introduce an RL-based algorithm to address it.
\subsection{Problem Formulation}
Using (\ref{qoe}), we can define the overall QoE of all users as $\mathrm{QoE} \triangleq \sum_{t\in\mathcal{T}} \mathrm{QoE}(t)$ with $\mathrm{QoE}(t)\triangleq {\sum_{k\in\mathcal{K}}\mathrm{QoE}_k(t)}$.  Based on the mathematical model in (\ref{etakn})--(\ref{qoe}),  it is evident that $\mathrm{QoE}$ is directly influenced by $\mathbf{l}$ and indirectly affected by $\mathbf{w}$ and $\boldsymbol{\eta}$ through the constraints in (\ref{lkn}), (\ref{etakn}), (\ref{maxpower}), (\ref{sinr}), and (\ref{slot}). As such, our objective is to maximize $\mathrm{QoE}$ by jointly optimizing $\mathbf{w}$, $\mathbf{l}$, and $\boldsymbol{\eta}$ subject to the constraints in (\ref{lkn}), (\ref{etakn}), (\ref{maxpower}), (\ref{sinr}),  and (\ref{slot}). Mathematically, this objective can be achieved by solving the following problem, denoted by $\mathcal P1$.
\begin{align*}
(\mathcal P1):\quad&\underset{{\mathbf w},\boldsymbol{\mathbf{l}},{\boldsymbol{\eta}}}{\rm {max}} \quad \mathrm{QoE}\nonumber\\
{\rm s.t.}\ &(\ref{lkn}), (\ref{etakn}), (\ref{maxpower}), (\ref{sinr}),  (\ref{slot}).
\end{align*}

Problem $\mathcal P1$ presents as a non-convex MINLP problem arising from the interplay of the continuous variable ${\mathbf w}$, the integer variables $(\mathbf{l}, {\boldsymbol{\eta}})$, and the non-convex constraint in (\ref{sinr}). Moreover, even after relaxing $\mathbf{l}$ and $ {\boldsymbol{\eta}}$ to be continuous, the resultant problem is still non-convex. Consequently, solving problem $\mathcal P1$ is an exceptionally challenging task, which requires substantial computational resources to achieve a globally optimal solution. 

\subsection{Problem Solution}
Problem $\mathcal P1$ is non-convex but it is convex with respect to $\mathbf{w}$ if $\mathbf{l}$ and $\boldsymbol{\eta}$ are given. As such, we can obtain an optimal beamforming design by solving the following subproblem $\mathcal P2$.
\begin{align*}
(\mathcal P2):\quad&\underset{{\mathbf w}}{\rm {max}} \quad \mathrm{QoE} \nonumber\\
&{\rm s.t.}\ (\ref{maxpower}),\ (\ref{sinr}),\ (\ref{slot}).
\end{align*}
Since the objective function is independent of ${\mathbf w}$ and the constraints in (\ref{maxpower}),\ (\ref{sinr}),\ (\ref{slot}) are convex, $\mathcal P2$ is a convex feasibility problem and can be solved efficiently using the CVX solver \cite{boyd}. Let $\mathbf{w}^*\triangleq (\mathbf{w}^*(t))_{t\in  \mathcal{T}}$ denote an optimal solution of $\mathcal P2$, where $\mathbf{w}^*(t)$ is the optimal solution in each time slot. We can find that time $t$ is separable from this problem. Due to the problem's time-separable structure, we can simplify the solving process by iteratively determining $\mathbf{w}^*(t)$ for each time slot within set $\mathcal{T}$. {The complexity of solving $\mathcal P2$ depends on the specific algorithm used. For example, if using an interior point method to solve $\mathcal P2$, the complexity can be approximately expressed as
$O\left(\sqrt{K M I + M} \right)$ \cite{boyd}}.

Given $\mathbf{w}^*$, $\mathcal P1$ is reduced to the following problem:  
\begin{align*}
(\mathcal P3):\quad &\underset{{\mathbf{l}},{\boldsymbol{\eta}}}{\rm {max}}\quad \mathrm{QoE}\nonumber\\
&{\rm s.t.}\ (\ref{lkn}),\ (\ref{etakn}),\ (\ref{slot}).\nonumber
\end{align*}Note that $\mathcal P3$ presents a challenge as an integer programming problem, which makes it quite difficult to obtain its optimal solution through traditional numerical optimization methods. In the following, we pivot toward employing an RL algorithm, i.e., the PPO algorithm, to solve $\mathcal P3$. 
{PPO is a type of actor-critic (AC) RL algorithm, particularly designed to handle either discrete variables or continuous variables individually\cite{ppo}.} Note that, in contrast other AC algorithms, PPO can  avoid divergence by incorporating proximal optimization methods and clipping mechanisms. Furthermore, PPO exhibits reduced sensitivity to hyperparameters, which improves the stability and convergence of the learning process. {To apply the PPO algorithm to solve $\mathcal P3$, we construct a finite  Markov Decision Process \cite{vtc} and consider ECS as the agent that interacts with the environment and determines $\mathbf{l}(t)$ and $\boldsymbol{\eta}(t)$}. At each time step, ECS observes the state $\mathcal{S}(t)$, takes action $\mathcal{A}(t)$ to obtain $\{\mathbf{l}(t),\boldsymbol{\eta}(t)\}$, and receives the reward $R(t)$. 
Specifically, $\mathcal{S}(t)$, $\mathcal{A}(t)$ and $R(t)$ in each training episode are defined as follows:
\begin{itemize}
\item{State: $\mathcal{S}(t)\triangleq \{r_k(t), Q_k(t-1), \tau, C_{\rm max}, {V_k(t)}\}_{k \in \mathcal{K}}$, which includes the transmission rate, the video's playback quality at last time slot $t-1$, delay constraint time, the computational capacity user device and the tile’s information $\mathcal{V}_k(t)\triangleq\{{\rm dist}_{k,n}(t), O_{k,n}(t)\}_{n\in \mathcal{N}_k(t)}$ in user's FoV. Note that, $\tau$ and $C_{\rm max}$ take different values in each episode, aiming to let the agent learn the optimal strategy under different conditions.} 
\item {Action: Based on $\mathcal{S}(t)$, ECS decides the bitrate selections and compression instructions of the requested tiles for all users at time $t$, i.e., $\mathcal{A}(t)\triangleq \{{\mathbf{l}}(t),{\boldsymbol{\eta}}(t)\}$.}
\item {Reward: It is designed as $R(t)\triangleq{\rm QoE}(t)$. Note that if the constraint in (\ref{slot}) cannot be satisfied, $R(t)$ is set to 0.}
\end{itemize}

The critic network is indicated by $V_\pi(\mathcal{S}(t))$, which is the state of the value function and defined as:
\begin{align}
    V_\pi(\mathcal{S}(t))= \mathbb{E}_{\mathcal{A}(t)\sim \pi} [G(t)\mid \mathcal{S}(t)].
\end{align}
$G(t)$ is the accumulated sum of discounted future reward $\sum_{i=t}^T\boldsymbol{\gamma}^{i-t}R(i)$, where $\gamma \in [0,1]$ is the discount factor that balances the importance between immediate and future rewards. {$V_\pi(\mathcal{S}(t)) $ represents the expected value of accumulated sum of discounted reward and PPO algorithm seeks the optimal action strategy by maximizing \(V_\pi(\mathcal{S}(t)) \).} The actor network, represented by $\pi_\theta(\mathcal{A}(t)\mid \mathcal{S}(t))$ with its parameter $\theta$, outputs the probability distribution of each action taken at a given state $\mathcal{S}(t)$. The formula for updating the learning rate denoted by $L^{\rm clip}\left(\theta\right)$, is defined as:
\begin{align}\label{clip}
    L^{\rm clip}\left(\theta\right)=\begin{cases}\left(1-\epsilon\right)D,&r\left(\theta\right)\leq1- \epsilon , \ D<0,\\\left(1+\epsilon\right)D,&r\left(\theta\right)\geq1+\epsilon \ ,\ D>0,\\r(\theta)D,&{\rm otherwise}.\end{cases}
\end{align} 
Here, $D$ is the advantage function to represent the difference between $\mathcal{A}(t)$ and the average value of all actions, $\epsilon$ is a clipped probability ratio used to limit the updating amplitude of the learning rate and $r(\theta)$ is the ratio between new and old policies.
{In each episode, PPO calculates the advantage function $D$ through the reward and the state value function, subsequently computes $r(\theta)$ and updates the actor network parameters through the PPO-clipping mechanism (\ref{clip}).} This limits the adjustment of the learning rate within a specified range, leading to improved stability.

Subsequently, the bitrate levels and compression indications are supplied as input to the CVX solver, facilitating the derivation of the beamformer ${\mathbf w}(t)$. Then, soft updates are performed to stabilize the actor and critic network during training. Consequently, in each episode, the current solution of the optimization variables $\{{\mathbf{l}},{\boldsymbol{\eta}}\}$ for problem $\mathcal{P}3$ are determined. The entire training process is depicted in Algorithm \ref{alg:alg1}.

\renewcommand{\algorithmicrequire}{ {\textbf{Initialization:}}}
\renewcommand{\algorithmicensure}{{ \textbf{Iteration:}}}
\renewcommand{\algorithmicprint}{ {\textbf{Initialization:}}}
\newcommand{\ILL}{\textbf{Initialize:}}
\begin{algorithm}[H]
\caption{{Proposed Algorithm for Joint Beamforming and Bitrate Selection.}}\label{alg:alg1}
\begin{algorithmic}[1]
{
\REQUIRE  {Initialize the environment and the parameters of actor and critic networks.}
\ENSURE
\FOR {episode = 0, 1, ...,} 
\FOR {$t$ = 1, 2, ...,$T$}
    \STATE Observe the environment and get the state $\mathcal{S}(t)$.
    \STATE Obtain action $\mathcal{A}(t)$ according to actor network.  
    \STATE Obtain beamformer ${\mathbf w}(t)$ by solving $\mathcal P2$ via CVX.
    \STATE Obtain the immediate reward 
    
    $R(t)=\begin{cases}
  {\rm QoE}(t),& \text{ if } \mathcal P2  \text{ is feasible}, \\
  0,& \text{ otherwise. }
    \end{cases}$
    \STATE Observe $\mathcal{S}(t+1)$ and store the tuple $(\mathcal{S}(t), \mathcal{A}(t), R(t), \mathcal{S}(t+1))$ in the replay memory.
    \ENDFOR
\IF{The length of the replay memory is equal to the batch size}
  \STATE Update the critic network and the actor network.
  \ENDIF
\ENDFOR
}
\end{algorithmic}
\label{alg1}
\end{algorithm}

\section{Simulation Results}

\begin{figure*}[h]
  \begin{minipage}{0.24\textwidth}
    \centering
    \includegraphics[width=\linewidth]{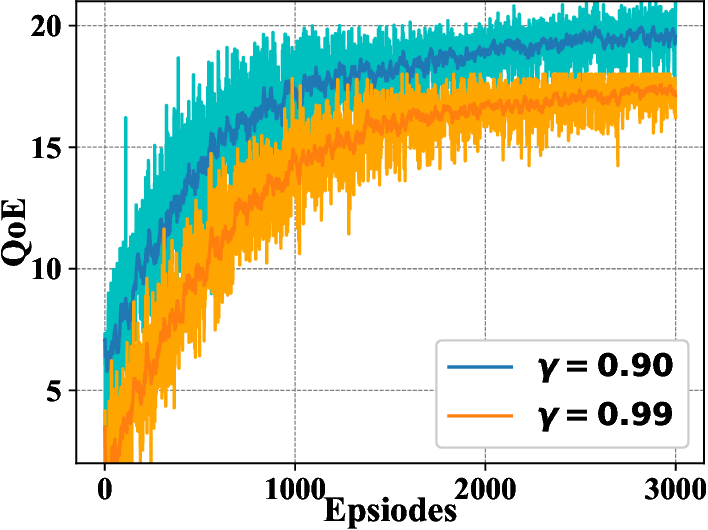}
    \caption{{Convergence of Algorithm~1.}}
    \label{convergence}
  \end{minipage}
  \hfill
    \begin{minipage}{0.24\textwidth}
    \centering
    \includegraphics[width=\linewidth]{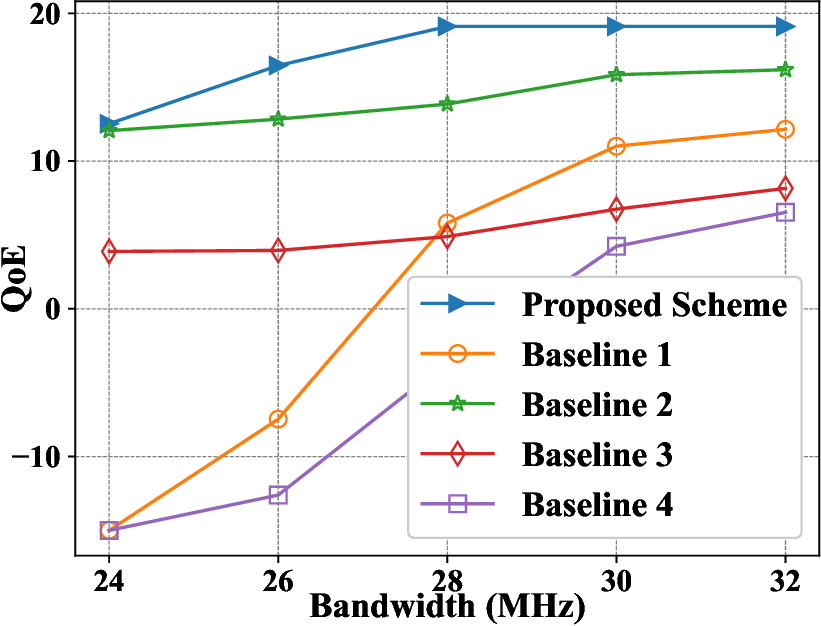}
    \caption{QoE versus $W$.}
    \label{bw}
  \end{minipage}
  \hfill
  \begin{minipage}{0.24\textwidth}
    \centering
    \includegraphics[width=\linewidth]{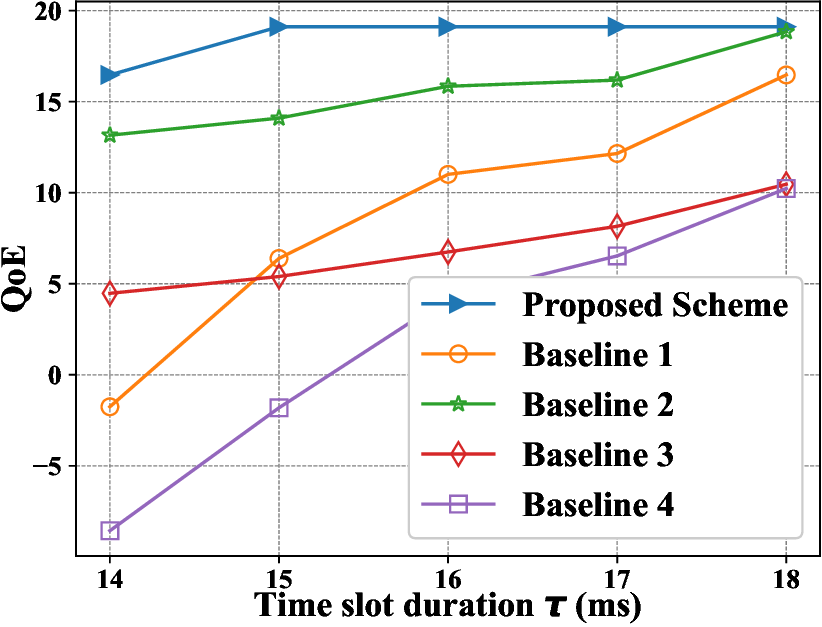}
    \caption{QoE versus $\tau$.}
    \label{tao}
  \end{minipage}
  \hfill
  \begin{minipage}{0.24\textwidth}
    \centering
    \includegraphics[width=\linewidth]{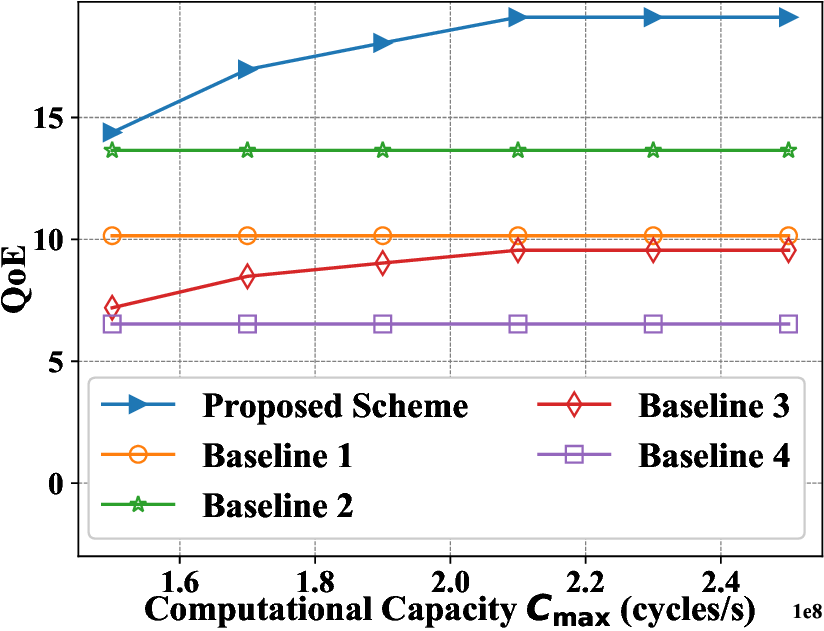}
    \caption{QoE versus $C_{\rm max}$.}
    \label{f}
  \end{minipage}
\end{figure*}

In this section, we evaluate the performance of our proposed PPO-based scheme. The detailed simulation settings are as follows. We consider a scenario with $M=4$ APs and $K=3$ users \cite{wenjoint}. For video parameters, video is spatially partitioned into $N=15$ tiles and there is $\mathcal{T}_k=6$ tiles in user's FoV \cite{vtc}. For all $m \in \mathcal{M}$, we use $W=28$ MHz, the maximum transmit power $P_m=38$ dBm, SINR threshold  $\xi=0.80$, time slot duration $\tau=15$ms, the compression ratio  $\varphi=0.80$ and weight factors $\alpha_1=0.5$, $\alpha_2=0.5$ \cite{mec,joint_ICC}. For comparison, we consider the following four baseline schemes: 
\begin{itemize}
\item{Baseline 1: This scheme employs cooperative transmission, and the bitrate and compression indication are fixed with ${l}_{k,n}(t)=2$ and ${{\eta}_{k,n}(t)=0}$, for all $n\in \mathcal{N}_k(t)$, $k \in \mathcal{K}$, and $t \in \mathcal{T}$. Then, the optimal beamforming design is determined by solving $\mathcal{P}2$ using CVX.}
\item{Baseline 2: This scheme utilizes cooperative transmission, while only transmits uncompressed tiles \cite{joint_ICC}, i.e., ${\eta}_{k,n}(t)=0$ for all ${n\in \mathcal{N}_k(t),k \in \mathcal{K}}$, and $t \in \mathcal{T}$. To obtain $\mathbf{l}$, we only need to change the action of PPO to $\mathcal{A}(t) = \{\mathbf{l}(t)\}$. And ${\mathbf{w}}$ is determined via CVX.}
\item{Baseline 3: This scheme employs non-cooperative transmission \cite{vtc}, i.e., $M=1$, and the beamforming vector ${\mathbf{w}}$ is determined through CVX, while ${\mathbf{l}}$ and ${\boldsymbol{\eta}}$ are obtained by the PPO algorithm.}
\item{Baseline 4: This scheme adopts non-cooperative transmission. Additionally, it opts for the same fixed uncompressed video bitrate as used in Baseline 1.}
\end{itemize}
{Fig.~\ref{convergence} shows the convergence behavior of our proposed Algorithm~1 under different discount factors. From Fig.~\ref{convergence}, we see that a lower discount factor ($\gamma = 0.90$) yields a higher QoE than a higher discount factor ($\gamma = 0.99$). This observation could be due to a higher $\gamma$ leading the agent to weigh future rewards more heavily, potentially at the expense of immediate rewards. Such prioritization could lead to suboptimal immediate decisions, resulting in a less effective action strategy in the short term. Note that the observation from Fig.~2 does not imply that a lower discount factor is universally superior. The choice of $\gamma$ is a hyperparameter decision, and different values of $\gamma$ will guide the agent toward different learning strategies. The optimal $\gamma$ is highly dependent on the specific characteristics of the environment and should be empirically determined through simulation experiments.}

Figs.~\ref{bw}--\ref{f} depict the relationship between QoE and system bandwidth $W$, time slot duration $\tau$, and computational capacity $C_{\rm max}$. Observations from Fig.~\ref{bw} and Fig.~\ref{tao} indicate that an increase in $W$ and $\tau$ corresponds to an increase in QoE for each scheme, which can be attributed to the expanded bandwidth and extended time resources, respectively. From Fig.~\ref{f}, it is evident that the QoE of our scheme and that of Baseline 3 escalates with the growth in computational capacity, a trend not observed in other schemes. This can be explained by the efficient transmission of compressed video tiles that reduces the load on wireless transmission by effectively utilizing the computational resources of users. Further insights gleaned from Figs.~\ref{bw}--\ref{f} reveal that Baseline 1 outperforms Baseline 4, underscoring the effectiveness of cooperative transmission techniques. Additionally, Baseline 3 exhibits superior performance over Baseline 4, which underscores the benefits of adaptive bitrate transmission. In particular, our proposed scheme demonstrates a marked improvement in QoE over both Baseline 1 and Baseline 3, indicating that the integration of cooperative transmission with bitrate adaptation significantly enhances the user experience.

\section{Conclusion}

In this work, we introduced a holographic video cooperative streaming framework designed for a generic wireless network where multiple APs can cooperatively transmit video with different bitrates to multiple users. We modeled a novel QoE metric specifically tailored for holographic video streaming, which can effectively encapsulate the nuances of holographic video quality, quality fluctuations, and rebuffering occurrences simultaneously. Based on the modeled QoE metric, we formulated a formidable QoE maximization problem and proposed a joint beamforming and bitrate control scheme using the PPO algorithm. The numerical results demonstrated the superiority of our scheme over representative baselines. {Future work may leverage artificial intelligence/machine learning techniques for advanced interference management and coordination in cooperative holographic video streaming.}

\bibliographystyle{IEEEtran}
\bibliography{IEEEabrv,QoE_Optimization_for_Holographic_Video_Streaming.bib}



\end{document}